\documentclass[11pt,a4paper]{JHEP3}
\usepackage{graphicx}

\title{Threshold production of unstable top}

\preprint{ALBERTA THY 11-11, TTK-11-44, SFB/CPP-11-52}

\author{Alexander A. Penin\\
  Department of Physics, University of Alberta,\\
  Edmonton AB T6G 2J1, Canada\\
  Institut f{\"u}r Theoretische Teilchenphysik, Karlsruhe
  Institute of Technology (KIT),\\
  76128 Karlsruhe, Germany\\
  Institute for Nuclear Research, Russian Academy of Sciences,\\
  119899 Moscow, Russia\\
  E-Mail: \email{penin@ualberta.ca}}
\author{Jan H. Piclum\\
  Institut f{\"u}r Theoretische Teilchenphysik und Kosmologie,
  RWTH Aachen,\\ 52056 Aachen, Germany\\
  E-Mail: \email{piclum@physik.rwth-aachen.de}}

\abstract{We develop a systematic approach to describe the finite lifetime
effects in the threshold production of top quark-antiquark pairs. It is based on
the nonrelativistic effective field theory with an additional scale $\rho^{1/2}
m_t$ characterizing the dynamics of the top-quark decay, which involves a new
expansion parameter $\rho=1-m_W/m_t$. Our method naturally resolves the problem
of spurious divergences in the analysis of the unstable top production. Within
this framework we compute the next-to-leading nonresonant contribution to the
total cross section of the top quark-antiquark threshold production in
electron-positron annihilation through high-order expansion in $\rho$ and
confirm the recently obtained result. We extend the analysis to the
next-to-next-to-leading ${\cal O}(\alpha_s)$ nonresonant contribution which is
derived in the leading order in $\rho$. The dominant nonresonant contribution to
the top-antitop threshold production in hadronic collisions is also obtained.}


\newcommand{\iep}{i\varepsilon}
\newcommand{\qe}{Q_e}
\newcommand{\qt}{Q_t}
\newcommand{\qb}{Q_b}
\renewcommand{\ae}{a_e}
\newcommand{\ve}{v_e}
\newcommand{\at}{a_t}
\newcommand{\vt}{v_t}
\newcommand{\ab}{a_b}
\newcommand{\vb}{v_b}
\newcommand{\cw}{c_w}
\newcommand{\sw}{s_w}
\newcommand{\bfm}[1]{\mbox{\boldmath$#1$}}
\newcommand{\bff}[1]{\mbox{\scriptsize\boldmath${#1}$}}

\begin{document}
\section{Introduction}
The threshold production of top quark-antiquark pairs at a future linear
collider may provide us with the most accurate information on the top-quark mass
and couplings crucial for our understanding of electroweak symmetry breaking and
mass generation mechanism~\cite{Martinez:2002st}. Due to renormalization group
suppression of the strong coupling  the nonrelativistic top-antitop pair is the
cleanest quarkonium-like  system. Its theoretical description is entirely based
on the first principles of QCD and is an ideal laboratory to develop the
nonrelativistic effective field theory approach. It is not surprising that since
the pioneering papers~\cite{Fadin:1987wz,Fadin:1988fn,Strassler:1990nw} the
top-antitop threshold production remaines in the focus of theoretical research
for  over two decades. A significant progress has been achieved in the analysis
of the higher order perturbative and relativistic corrections in the strong
coupling constant $\alpha_s$ and the heavy-quark velocity $v$. Sizable
next-to-next-to-leading order (NNLO) corrections to the total cross section have
been found by several groups
\cite{Hoang:1998xf,Melnikov:1998pr,Beneke:1999qg,Penin:1998mx,Nagano:1999nw}
that stimulated the study of the higher orders of perturbation theory. Currently
a bulk of the third order corrections is available
\cite{Kniehl:1999mx,Kniehl:1999ud,Kniehl:2002br,Penin:2002zv,Kniehl:2002yv,
Hoang:2003ns,Penin:2005eu,Beneke:2005hg,Marquard:2006qi,Beneke:2007gj,
Beneke:2007pj,Beneke:2008cr,Marquard:2009bj, Anzai:2009tm,Smirnov:2009fh} with
only a few Wilson coefficients still missing, and the N$^3$LO analysis is likely
to be completed in the foreseeable future.  Moreover the  higher order
logarithmically enhanced corrections have been resummed  through the effective
theory renormalization group methods
\cite{Hoang:2000ib,Penin:2004ay,Pineda:2006ri}.

At the same time much less attention has been paid to the analysis of the
effects related to the instability of the top quark
\cite{Bigi:1991mi,Melnikov:1993np,Penin:1998ik,Hoang:2004tg}. The width of the
electroweak top-quark  decay $t\to W^+b$, $\Gamma_t\approx 1.5$~GeV, is
comparable to the binding energy of a would-be toponium ground state and has a
dramatic effect on the threshold production. It serves as an infrared cutoff,
which makes the process perturbative in the whole threshold region, and smears
out all the Coulomb-like resonances below the threshold leaving  a single well
pronounced peak in the cross section. The standard prescription in the analysis
of the unstable top-quark production consists of the complex shift  $E\to
E+i\Gamma_t$, where $E$ is the top-antitop pair energy counted from the
threshold  \cite{Fadin:1987wz}.  Though this procedure incorporates  the
dominant effect of the finite top-quark width, it does not fully account for
nonresonant processes like $e^+e^-\to t W^-\bar{b}$, $e^+e^-\to b W^+\bar{t}$ or
$e^+e^-\to W^+W^- b\bar{b}$ where the intermediate top quark is not on its
(complex) mass shell.  Such processes {\em cannot be distinguished} from the
resonant $t\bar{t}$ production, which has the same final states due to the
top-quark instability. Moreover, a na\"ive  use of the above prescription
results in 
spurious divergences of the cross section in the nonrelativistic effective field
theory \cite{Bigi:1991mi,Penin:1998ik} and it has to be elaborated to make the
high-order calculations self-consistent. Recently  two  different approaches
have been used to refine the analysis of  the finite width effect. The first is
based on so-called ``phase space matching'' \cite{Hoang:2010gu} while the second
\cite{Beneke:2010mp} relies on  the effective theory of unstable particles
\cite{Beneke:2003xh}. In particular, in ref.~\cite{Beneke:2010mp} the NLO
nonresonant contribution to the total cross section has been computed.

In the present paper we develop an alternative approach which  systematically
takes into account the  effect of top-quark instability. The approach is
applicable to the threshold production of an unstable particle which is almost
degenerate in mass with one of its decay products. It  introduces into the
nonrelativistic effective field theory \cite{Caswell:1985ui,Bodwin:1994jh} an
additional scale $\rho^{1/2} m_t$ characterizing the dynamics of the top-quark
decay into a nonrelativistic $W$-boson and an ultrarelativistic  bottom quark.
The parameter $\rho$ is related to the difference of the top quark and $W$-boson
masses, $\rho=1-m_W/m_t$, and $\rho^{1/2}$ plays the role of the heavy quark
velocity in standard potential nonrelativistic QCD (pNRQCD)
\cite{Pineda:1997bj,Brambilla:1999xf}. The new scale obeys the hierarchy
$vm_t\ll\rho^{1/2} m_t\ll m_t$ or $v\ll \rho^{1/2} \ll1$  and the cross section
is constructed as  a series in the scale ratios. The method is applicable for
the analysis of the high order corrections and naturally resolves the problem of
the spurious divergences.

In the next section we outline the main concept and derive the NLO nonresonant
contribution to the total cross section in the leading order in $\rho$. In
section~\ref{sec::rhoexp} we describe the calculation of the high order terms of
the expansion. The strong coupling corrections to the result of
section~\ref{sec::main}  are computed in section~\ref{sec::alexp}. In
section~\ref{sec::log} we show how the spurious effective theory divergences
associated with the top-quark instability are eliminated within our approach.
Section~\ref{sec::res} collects the final result and  numerical estimates.
Application of our result to the analysis of the experimentally measured
cross section with invariant mass cuts is discussed in section \ref{sec::cut}.
Section~\ref{sec::con} is our conclusion.
\section{Finite width effect beyond the complex energy shift}
\label{sec::main}
\begin{figure}[t]
  \begin{center}
    \includegraphics[width=0.4\textwidth]{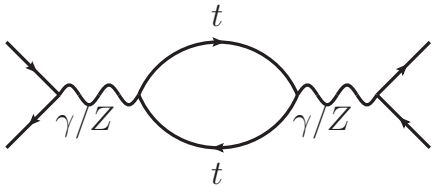}
  \end{center}
  \caption{\label{fig::lo}$e^+e^-$ forward scattering diagram corresponding to
  the leading order top-quark pair production process.}
\end{figure}
\begin{figure}[t]
  \begin{center}
    \includegraphics[width=\textwidth]{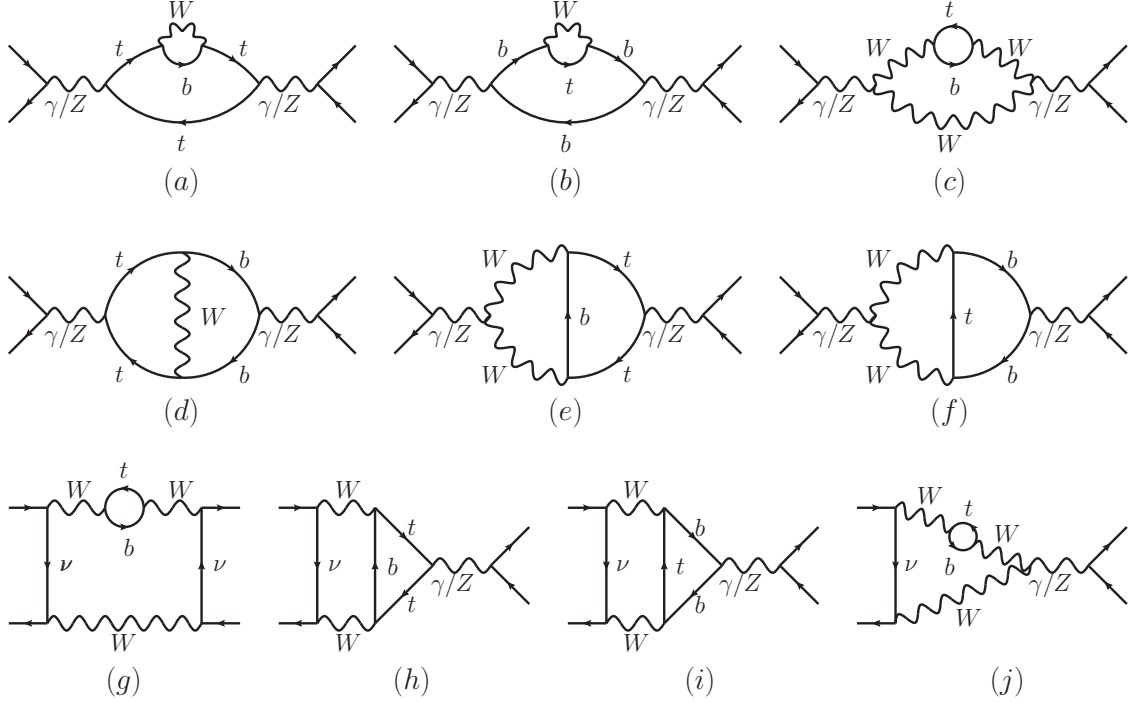}
  \end{center}
  \caption{\label{fig::dias}$e^+e^-$ forward scattering diagrams
    containing $b W^+\bar{t}$ and $t W^-\bar{b}$ cuts.}
\end{figure}

In the Born approximation the total cross section of top-antitop production
in electron-positron annihilation is related through the optical theorem to the
imaginary part of the one-loop forward scattering amplitude shown in
figure~\ref{fig::lo}. The corresponding expression for the normalized cross
section $R ={\sigma(e^+e^-\to t\bar t)/\sigma_0}$, $\sigma_0=4\pi\alpha^2/(3s)$,
in the threshold region $s\approx 4m_t^2$ can be obtained by the standard
nonrelativistic expansion of the top-quark vertices and propagators in $v$
and reads
\begin{eqnarray}
R^{Born}_{res}= \left[ \qe^2\qt^2 + \frac{2\qe\qt\ve\vt}{1-x_Z} +
\frac{(\ae^2+\ve^2)\vt^2}{(1-x_Z)^2} \right]{6\pi N_c\over m_t^2}\,
{\rm Im}[G_0(0,0,E+\iep)]
+\ldots\,,
\label{res}
\end{eqnarray}
where the ellipsis stands for the relativistic corrections, $Q_f$ is the
electric charge of fermion $f$ in units of the positron charge, $N_c=3$ is the
number of colors, and $x_Z=m_Z^2/(4m_t^2)$ with the $Z$-boson mass $m_Z$. The
couplings of fermion $f$ to the $Z$-boson are
\begin{equation}
v_f = \frac{I^3_{w,f} - 2\sw^2Q_f}{2\sw\cw}\,, \qquad a_f =
\frac{I^3_{w,f}}{2\sw\cw} \,,
\end{equation}
where $I^3_{w,f}$ is the third component of the fermion's weak isospin and $\sw$
($\cw$) is the sine (cosine) of the weak mixing angle. Note that only the vector
coupling of the top quark gives the leading order contribution and the axial
coupling is suppressed by an additional power of $v$. The last factor in
eq.~(\ref{res}) is
\begin{eqnarray}
G_0(0,0,E)=\int {{\rm
d}^{d-1}\bfm{p}\over (2\pi)^{d-1}}{m_t\over{\bfm{p}^2-m_tE}}
=-{m_t^2\over 4\pi}\sqrt{{-E\over m_t}}\,,
\label{g0}
\end{eqnarray}
which is nothing but the Green's function of the free Schr\"odinger equation at
the origin. Formally the integral in eq.~(\ref{g0}) is linearly divergent  but
the divergent part is real and does not contribute to the cross section. To
handle the divergence we use dimensional regularization with $d=4-2\epsilon$,
where the integral~(\ref{g0}) is finite even for $\epsilon =0$.
The strong interaction has a significant impact on the threshold cross section.
Close to threshold when $v\sim \alpha_s$ the Coulomb effects become
nonperturbative and have to be resummed to all orders in $\alpha_s$ by
substituting eq.~(\ref{g0}) with the full Coulomb Green's function
\begin{eqnarray}
G_C(0,0;E)&=&G_0(0,0;E)+G_1(0,0;E)-{C_F\alpha_sm_t^2\over
4\pi}\left[\Psi\left(1- {C_F\alpha_s\over
2}\sqrt{m_t\over -E} \right)+\gamma_E\right]\,,
\nonumber
\\
&&\label{gc}
\end{eqnarray}
where $\Psi$ is the logarithmic derivative of the Gamma function and
$C_F=4/3$. The one-gluon exchange contribution $G_1(0,0;E)$ is ultraviolet
divergent. Again for stable top quarks the divergent part is real and does not
contribute to eq.~(\ref{res}). In the $\overline{\rm MS}$ subtraction scheme
this term reads
\begin{eqnarray}
G_1(0,0;E)&=&-{C_F\alpha_sm_t^2\over
8\pi}\left[\ln\left({-m_tE\over \mu^2}\right)-1+2\ln{2}\right]\,.
\label{g1}
\end{eqnarray}
Let us now consider the top-quark decay. Every decay process is suppressed by
the electroweak coupling constant $\alpha_{ew}$. We adopt the standard power
counting rules $\alpha_s\sim v$, $\alpha_{ew}\sim v^2$. Thus to NNLO if the
top quark decays the antiquark may be treated as a stable particle and vice
versa.  The dominant effect of the top-quark instability is related to the
imaginary part of its mass operator in diagram~\ref{fig::dias}$(a)$.
In the massless bottom quark approximation and with the off-shell momentum $p$
the mass operator reads
\begin{equation}
 {\rm Im}[\Sigma^{(0)}(p^2)]={G_F\over
16\pi\sqrt{2}}\,{p^3}\left(1+2{m^2_W\over p^2}\right)\left(1-{m^2_W\over
p^2}\right)^2
\theta(p^2-m_W^2)\,,
\label{sigma0}
\end{equation}
where $G_F$ is the Fermi constant and we use the approximation $V_{tb}= 1$.
Close to the mass shell one has $p^2=m_t^2-2(\bfm{p}^2-m_tE)+\ldots$, where
$\bfm{p}$ is the spatial momentum of the top quark, and the mass operator can be
expanded in $z=(\bfm{p}^2-m_tE)/m_t^2\ll 1$
\begin{eqnarray}
 {\rm
Im}[\Sigma^{(0)}(z)]&=&{\Gamma_t\over 2}\left(1-{4z\over
(1-x^2)}+{4z^2\over
(1-x^2)^2}\right)\theta(1-x^2-2z)+\ldots
\nonumber
\\
&=& {\Gamma_t\over 2}-{\Gamma_t\over 2}\!\left[\theta(x^2+2z-1)+\left({4z\over
(1-x^2)}-{4z^2\over
(1-x^2)^2}\right)\theta(1-x^2-2z)\right]\!+\!\ldots\,.
\nonumber
\\
\label{sigma0os}
\end{eqnarray}
where $x=m_W/m_t$, $\Gamma_t=\Gamma^{(0)}_t+{\cal O}(\alpha_s)$ and
\begin{equation}
\Gamma^{(0)}_t={G_Fm_t^3\over 8\pi\sqrt{2}}(1+2x^2)(1-x^2)^2\,,
\label{gamma0}
\end{equation}
is the leading order top-quark electroweak width. Note that in the
expansion~(\ref{sigma0os}) we consider $1-x=\rho$ to be of the same order of
magnitude as $z$. The first term in the last line of eq.~(\ref{sigma0os})
describes the standard shift of the pole position of the top quark  propagator
into the unphysical sheet of the complex energy plane characteristic for
unstable particles. After Dyson resummation it replaces the argument of
eq.~(\ref{g0}) by $E+i\Gamma_t$, which is the original prescription of
ref.~\cite{Fadin:1987wz}. Eq.~(\ref{sigma0os}), however, has the remainder which
also contributes to the imaginary part of the forward scattering amplitude.
Since the remainder vanishes for on-shell top quark, it represents the
nonresonant process $e^+e^-\to b W^+\bar{t}$ or $e^+e^-\to t W^-\bar{b}$.

In the nonresonant contribution the integral over the virtual momentum $\bfm{p}$
is saturated by the region $|\bfm{p}|\sim \rho^{1/2} m_t$. The main idea of our
approach is that if $\rho$ is considered as a small parameter this momentum
region corresponds to a nonrelativistic top quark with the energy $p_0-m_t\sim
\bfm{p}^2/m_t \sim \rho m_t$,  and one may apply the well elaborated pNRQCD
tools for the calculation of the cross section. In this complementary
nonrelativistic expansion the  heavy quark velocity $v$ is replaced by
$\rho^{1/2}$ and we have the hard scale $m_t$, the soft scale  $\rho^{1/2}m_t$,
and the ultrasoft scale $\rho m_t$.  Note that in this case the $W$-boson is
also nonrelativistic while the bottom quark is ultrarelativistic with the
ultrasoft four-momentum of order $\rho m_t\gg m_b$. Further expansion of
eq.~(\ref{sigma0os}) in $\rho\sim z$ gives
\begin{eqnarray}
{\rm Im}[\Sigma^{(0)}(z)]&=& {\Gamma_t\over 2}-{\Gamma_t\over 2}\left[
\theta(z-\rho)+\left({2z\over\rho}-{z^2\over \rho^2}\right)\theta(\rho-z)
+{\cal O}(\rho,z)\right]\,.
\label{sigma0exp}
\end{eqnarray}
By inserting  the second term of eq.~(\ref{sigma0exp}) into the
diagram~\ref{fig::dias}$(a)$ one obtains the following contribution to the
nonresonant cross section
\begin{eqnarray}
R_1= \left[ \qe^2\qt^2 + \frac{2\qe\qt\ve\vt}{1-x_Z}+\frac{(\ae^2+\ve^2)\vt^2}
{(1-x_Z)^2}\right]{ N_c \Gamma_t\over m_t}\delta_1(1+{\cal O}(\rho))\,,
\label{r1}
\end{eqnarray}
where   $\delta_1=\delta^{(0)}_1+{\cal O}(\alpha_s)$. The leading order result
reads
\begin{eqnarray}
\delta^{(0)}_1&=&-6\pi\left[\int {{\rm d}^{3}\bfm{p}\over
(2\pi)^{3}}\theta({\bfm{p}^2-\rho
m_t^2}){m_t\over{\bfm{p}^4}}
+\int {{\rm d}^{3}\bfm{p}\over (2\pi)^{3}}\theta({\rho m_t^2-\bfm{p}^2})
\left({2\over\rho}{\bfm{p}^2\over m_t^2}-{1\over\rho^2}{\bfm{p}^4\over
m_t^4}\right){m_t\over{\bfm{p}^4}}\right] \nonumber \\
&=&-{8\over \pi}{1\over \rho^{1/2}}\,.
\label{d10}
\end{eqnarray}
In eq.~(\ref{d10}) we neglect the $m_tE$ term in $z$ since it gives a subleading
contribution suppressed by $m_tE/\bfm{p}^2\sim v^2/\rho\ll 1$ according to our
scale  hierarchy.  Note that one can obtain the result ~(\ref{d10}) without
subtracting the first term of  eq.~(\ref{sigma0exp}) by direct evaluation of the
corresponding Feynman integral in dimensional regularization. In this case the
resonance contribution vanishes since eq.~(\ref{g0}) becomes a scaleless
integral.

Eq.~(\ref{r1}) however does not give the full nonresonant contribution and one
has to take into account all diagrams of this order with the $b W^+\bar{t}$
or $t W^-\bar{b}$ cuts given in figure~\ref{fig::dias}. The calculation is
significantly simplified within the nonrelativistic effective theory where all
the propagators which are off-shell by the amount $m_t$ collapse, giving rise
to new effective theory vertices. From the practical point of view, however, it
is more convenient to directly expand the full theory Feynman integrals in
$\rho$ than to use the effective theory Feynman rules.  We found that beside
diagram~\ref{fig::dias}$(a)$ only diagram~\ref{fig::dias}$(g)$
gives a leading order contribution in  $\rho$. After the expansion the off-shell
neutrino  propagators shrink to points and the diagram becomes similar to
figure~\ref{fig::dias}$(a)$. In this case the contribution comes from the
imaginary part of the vacuum polarization operator of the heavy-light vector
current correlator
\begin{equation}
{\rm Im}[\Pi^{(0)}(p^2)]={N_c\over 24\pi}\left(2+{m^2_t\over p^2}\right)\left(1-
{m^2_t\over p^2}\right)^2\theta(p^2-m_t^2)\,,
\label{pi0}
\end{equation}
where $p^2=2m_t^2-m_W^2-2(\bfm{p}^2-m_tE)+\ldots$. It has the following
expansion
\begin{eqnarray}
{\rm Im}[\Pi^{(0)}(z)]&=& {N_c\rho^2\over
2\pi}\left[\left(1-{z\over\rho}\right)^2
\theta(\rho- z)+{\cal O}(\rho,z)\right]\,.
\label{pi0exp}
\end{eqnarray}
The corresponding  contribution to the nonresonant cross section reads
\begin{eqnarray}
R_2= {1\over \sw^4} { N_c\Gamma_t\over m_t}\delta_2(1+{\cal O}(\rho))\,,
\label{r2}
\end{eqnarray}
where
\begin{eqnarray}
\delta^{(0)}_2&=&2\pi
\int {{\rm d}^{3}\bfm{p}\over (2\pi)^{3}}\theta({\rho m_t^2-\bfm{p}^2})\left(
1-{\bfm{p}^2\over \rho m_t^2}\right)^2{m_t\over (2\rho m_t^2-\bfm{p}^2)^2}
\nonumber\\
&=&
\left({17\over 6}-{9\sqrt{2}\over 4}\ln\left(1+\sqrt{2}\right)\right){1\over
  \pi}{1\over \rho^{1/2}}\,.
\label{d20}
\end{eqnarray}
Each $W$-boson propagator brings a factor $(2\rho m_t^2-\bfm{p}^2)^{-1}$ to
eq.~(\ref{d20}) which is regular at  $\bfm{p}^2= 0$ so that the diagram does not
have a resonance contribution. Finally for the leading nonresonant contribution
to the cross section we get
\begin{eqnarray}
R_{nr}&=&- {8N_c\over \pi \rho^{1/2}}{\Gamma_t\over m_t} \Bigg[
\left(\qe^2\qt^2
+ \frac{2\qe\qt\ve\vt}{1-x_Z} +\frac{(\ae^2+\ve^2)\vt^2}{(1-x_Z)^2} \right)
\nonumber
\\
&&\left.
-{1\over \sw^4} \left({17\over 48}-{9\sqrt{2}\over
    32}\ln\left(1+\sqrt{2}\right)\right)+{\cal O}(\rho,\alpha_s)\right]\,.
\label{rnrlo}
\end{eqnarray}
Let us now compare our approach to the one of \cite{Beneke:2010mp}. In
ref.~\cite{Beneke:2010mp} the scales $m_t$ and  $\rho^{1/2} m_t$  are
considered to be of the same order and are integrated out simultaneously. As a
result the nonresonant contribution is represented by the imaginary part of the
Wilson coefficient of the local four-fermion $e^+e^-e^+e^-$ operators, i.e.
every  diagram in figure~\ref{fig::dias} shrinks to a point. The method
requires more complex calculations but gives the exact dependence of the cross
section on $\rho$. However, in the next section we show how to  compute a
sufficient number of terms of the expansion in $\rho$ to ensure very good
accuracy  of the approximation for the physical value $\rho\approx 0.53$. At the
same time application of the method  \cite{Beneke:2010mp} to the calculation of
the strong interaction corrections to the nonresonant contribution seems to
become technically complicated while our approach does not as we show in
section~\ref{sec::alexp}.
\section{\boldmath Relativistic corrections}
\label{sec::rhoexp}
To obtain the high order terms of the nonrelativistic expansion of the
nonresonant
cross section in $\rho$ we use the method of
regions~\cite{Beneke:1997zp,Smirnov:2002pj}. By
the optical theorem the problem is reduced to the calculation of the
contribution
from $b W^+\bar{t}$ and $t W^-\bar{b}$ cuts to the imaginary part of the
two-loop
electron-positron forward scattering amplitude, figure~\ref{fig::dias}. As an
example let us describe the evaluation of diagram~\ref{fig::dias}$(e)$. The
corresponding scalar integral reads
\begin{eqnarray}
  && \int \frac{{\rm d}^dk\, {\rm d}^dl}{(l^2+\iep) (k^2+q\cdot k+\iep)
    (k^2-q\cdot k+\iep) [(k+l)^2+q\cdot(k+l)+m_t^2\rho(2-\rho)+\iep]} \nonumber
  \\
  && \times
  \frac{1}{[(k+l)^2-q\cdot(k+l)+m_t^2\rho(2-\rho)+\iep]}\,, \label{eq::intfull}
\end{eqnarray}
where $q=(2m_t,\bfm{0})$ is the photon/$Z$-boson momentum corresponding to the
top-antitop threshold. In the limit $\rho\to0$ the only  nonvanishing
contribution to the imaginary part comes from the region of potential momentum
$k$ and ultrasoft momentum $l$
\begin{equation}
  k_0 \sim m_t\rho\,, \qquad \bfm{k} \sim m_t \rho^{1/2}\,, \qquad l \sim
  m_t\rho\,. \label{eq::region}
\end{equation}
By imposing  this scaling we expand eq.~(\ref{eq::intfull}) in $\rho$.
For example in the leading order we obtain
\begin{eqnarray}
  && \int \frac{{\rm d}k_0\, {\rm d}^{d-1}\bfm{k}\,{\rm d}l_0\, {\rm
      d}^{d-1}\bfm{l}}{(l^2+\iep) (-\bfm{k}^2+2m_tk_0+\iep)
    (-\bfm{k}^2-2m_tk_0+\iep) [-\bfm{k}^2+2m_t(k_0+l_0)+2m_t^2\rho+\iep]}
  \nonumber \\
  && \times
  \frac{1}{[-\bfm{k}^2-2m_t(k_0+l_0)+2m_t^2\rho+\iep]}\,. \label{eq::intexp}
\end{eqnarray}
Then we evaluate the integrals over the zero components of the loop momenta by
closing the integration contours in the upper half of the corresponding complex
planes. At this step we have to distinguish between the different cuts of the
diagram and pick up only those poles which correspond to $b W^+\bar{t}$ and $t
W^-\bar{b}$ cuts. The integral over $\bfm{k}$ becomes a one-loop massive tadpole
integral in $d-1$ dimensions and can be performed easily. Integration over
$\bfm{l}$ yields a Gauss hypergeometric function ${}_2$F$_1$ with half-integer
parameters, which can be expanded in $\epsilon$ with the {\tt Mathematica}
package {\tt HypExp}~{\tt 2}~\cite{Huber:2007dx}.

The apparently more complicated diagrams
\ref{fig::dias}$(g)$--\ref{fig::dias}$(j)$ with $t$-channel neutrino propagators
do not pose any new problems since the off-shell neutrino propagators do not
depend on the loop momentum after the expansion and the resulting integrals can
be computed in the same way. We use {\tt QGRAF}~\cite{Nogueira:1991ex}, {\tt
q2e}, and {\tt exp}~\cite{Harlander:1997zb,Seidensticker:1999bb} to generate the
diagrams and produce {\tt FORM}-readable expressions for the amplitudes. The
expansions and integrations are performed with custom code written in {\tt
FORM}~\cite{Vermaseren:2000nd}. The result of the calculation is presented in
section~\ref{sec::res}.
\begin{figure}[t]
  \begin{center}
    \includegraphics[width=\textwidth]{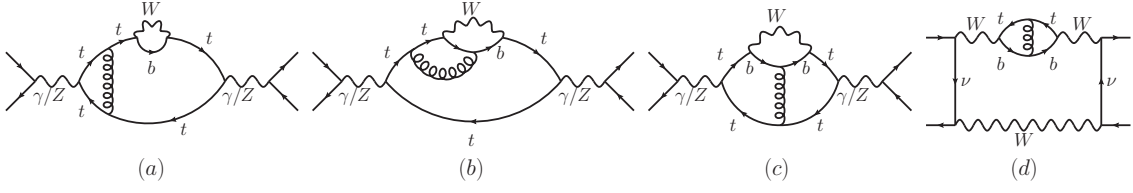}
  \end{center}
  \caption{\label{fig::ascorr}$e^+e^-$ forward scattering diagrams
    contributing to the $\mathcal{O}(\alpha_s)$ correction in the leading
    order in $\rho$. The curly lines denote gluons.}
\end{figure}
\section{Strong coupling corrections}
\label{sec::alexp}
In the leading order in $\rho$ the $\alpha_s$ corrections are obtained by the
gluon dressing of diagrams~\ref{fig::dias}$(a)$ and~\ref{fig::dias}$(g)$
shown in figure~\ref{fig::ascorr}.  Since the top quark is  nonrelativistic we
can apply the standard  pNRQCD arguments to identify the relevant regions of
virtual momentum. For diagram~\ref{fig::ascorr}$(a)$  the leading
contribution comes from the  Coulomb gluon with the potential momentum $q_0\sim
\bfm{q}^2/m_t\sim \rho m_t$. As in the case of the  Coulomb gluon exchange in
pNRQCD, the corresponding correction is  power enhanced due to the Coulomb
singularity but the enhancement factor here is $1/\rho^{1/2}$ rather than $1/v$.
An important difference with respect to pNRQCD is that since
$\alpha_s/\rho^{1/2}\ll 1$ one does not need to resum the Coulomb corrections to
all orders.  Following the analysis of section~\ref{sec::main} one gets the
potential Coulomb gluon contribution of the following form
\begin{eqnarray}
\left.\delta^{(1)}_{1a}\right|_{\rho^{-1}}&=&-12\pi{\rm Re}
\Bigg\{\int {{\rm d}^{3}\bfm{p}\over (2\pi)^{3}}{{\rm d}^{3}\bfm{p}'\over
(2\pi)^{3}}
{m_t\over{(\bfm{p}^2-m_t(E+i\Gamma_t))^2}}{m_t\over{\bfm{p'}^2-m_t(E+i\Gamma_t)}
}
{4\pi C_F\alpha_s\over (\bfm{p}-\bfm{p}')^2}
\nonumber
\\
&&\times\left[\theta({\bfm{p}^2-\rho m_t^2})+\theta({\rho m_t^2-\bfm{p}^2})
\left({2\over\rho}{\bfm{p}^2\over m_t^2}-{1\over\rho^2}{\bfm{p}^4\over
m_t^4}\right)\right]\Bigg\}
\nonumber
\\
&=&3\left(L_E +\frac{1}{2}+2\ln2\right){C_F\alpha_s\over\rho}\,,
\label{d1acoul}
\end{eqnarray}
where $L_E=\ln\left({\sqrt{E^2+\Gamma_t^2}\over \rho m_t}\right)$ and
$\delta^{(1)}_1$ defines the correction to eq.~(\ref{r1}),
$\delta_1=\delta^{(0)}_1+\delta^{(1)}_1+{\cal O}(\alpha^2_s)$. Note that one has
to keep a nonzero (complex) energy in the top-quark propagator in
eq.~(\ref{d1acoul}) since it serves as an infrared regulator. The contribution
of the hard gluon with momentum $q\sim m_t$ in
diagram~\ref{fig::ascorr}$(a)$ can be related to the Wilson coefficient in the
nonrelativistic expansion of the vector current $j_\mu$
\begin{equation}
{\bfm j}=c_v\psi^\dagger{\bfm \sigma}\chi+{d_v\over6 m^2}
\psi^\dagger{\bfm \sigma}\mbox{\bfm D}^2\chi+\ldots\,,
\label{currentexp}
\end{equation}
where  $\psi$ and $\chi$ are the nonrelativistic quark and antiquark
two-component Pauli spinors and the Wilson (matching) coefficients are
$c_v=1-2C_F\alpha_s/\pi+\ldots$ and $d_v=1+\ldots$.  The ${\cal O}(\alpha_s)$
term in $c_v$ results in the following correction to  $\delta_1$ in
eq.~(\ref{g0})
\begin{equation}
\left.\delta^{(1)}_{1a}\right|_{\rho^{-1/2}}=-{4C_F\alpha_s\over\pi}\delta^{(0)}
_1
={32}{C_F\alpha_s\over\pi^2}{1\over \rho^{1/2}}\,.
\label{d11ahard}
\end{equation}
Diagram~\ref{fig::ascorr}$(b)$ is determined by the correction to the mass
operator in the limit $z\to 0$ that can be read off the result of
ref.~\cite{Jezabek:1988iv}
\begin{eqnarray}
 {\rm Im}[\Sigma^{(1)}(z)]={C_F\alpha_s\over\pi}\left[{9\over 4}-{2\over 3}\pi^2
-{3\over 2}\ln(2\rho)-{3\over 2}\ln\left(1-{z\over \rho}\right)+{\cal
O}(\rho,z)\right]
{\rm Im}[\Sigma^{(0)}(z)]\,.
\label{sigma1}
\end{eqnarray}
The correction factor should in principle be substituted into the
integral~(\ref{d10}). However all the terms in eq.~(\ref{sigma1}) except the
last logarithmic one actually describe the corrections to the top-quark width
\cite{Jezabek:1988iv}, $\Gamma_t=\Gamma^{(0)}_t+\Gamma^{(1)}_t$, where
\begin{equation}
 \Gamma^{(1)}_t={C_F\alpha_s\over\pi}\left({9\over 4}-{2\over 3}\pi^2
-{3\over 2}\ln(2\rho)+{\cal O}(\rho)\right)\Gamma^{(0)}_t\,.
\label{gamma1}
\end{equation}
They all disappear when  the leading order top-quark width in eq.~(\ref{r1})
is replaced with its corrected (physical) value.  The remaining term gives
\begin{eqnarray}
\delta^{(1)}_{1b}&=&-{9}{C_F\alpha_s}\int {{\rm
d}^{3}\bfm{p}\over(2\pi)^{3}}
\theta({\rho m_t^2-\bfm{p}^2})\left(1-{\bfm{p}^2\over \rho m_t^2}\right)^2
\ln\left(1-{\bfm{p}^2\over \rho m_t^2}\right){m_t\over \bfm{p}^4}
\nonumber
\\
&=&-2\left({7}-12\ln{2}\right){C_F\alpha_s\over\pi^2}{1\over
\rho^{1/2}}\,.
\label{d11b}
\end{eqnarray}
The same procedure applies to diagram~\ref{fig::ascorr}$(d)$ which is
determined by the correction to the polarization function
\cite{Chetyrkin:2000mq}
\begin{eqnarray}
 {\rm Im}[\Pi^{(1)}(z)]={C_F\alpha_s\over\pi}\left[{9\over 4}+{1\over 3}\pi^2
-{3\over 2}\ln(2\rho)-{3\over 2}\ln\left(1-{z\over \rho}\right)+{\cal
O}(\rho,z)\right]
{\rm Im}[\Pi^{(0)}(z)]\,.
\label{pi1}
\end{eqnarray}
After factoring out the corrections to the top-quark width one gets
\begin{eqnarray}
\delta^{(1)}_{2}&=&{C_F\alpha_s}\left[\pi \delta^{(0)}_2 -{3}
\int {{\rm d}^{3}\bfm{p}\over (2\pi)^{3}}\theta({\rho m_t^2-\bfm{p}^2})\left(1
-{\bfm{p}^2\over \rho m_t^2}\right)^2\ln\left(1-{\bfm{p}^2\over \rho
m_t^2}\right)
{m_t\over (2\rho m_t^2-\bfm{p}^2)^2}\,\right]
\nonumber
\\
&=&\bigg[ \frac{22}{3} +
\frac{17\pi^2}{6} -
\frac{17}{2} \ln2 + \left(2 - 3\pi^2 + 9\ln2 \right)
\frac{3\sqrt{2}}{4}\ln\left(1+\sqrt{2}\right)
\nonumber
\\
&& -\frac{27\sqrt{2}}{8} \left(
\ln^2\left(1+\sqrt{2}\right)+\mathrm{Li}_2
\left(2\sqrt{2}-2\right)\right) \bigg]{C_F\alpha_s\over\pi^2}{1\over
\rho^{1/2}}\,,
\label{d12}
\end{eqnarray}
where $\mathrm{Li}_2$ stands for the dilogarithm function.

The potential gluon contribution to the nonfactorizable
diagram~\ref{fig::ascorr}$(c)$ in the leading order in $v$ vanishes for the
total cross section  \cite{Melnikov:1993np}. The hard gluon contribution to this
diagram is power suppressed. Thus the diagram vanishes in our approximation.
This completes the calculation of the dominant ${\cal O}(\alpha_s)$ corrections.
In the next section, however, we present the analysis of the corrections
suppressed by an additional power of $\rho^{1/2}$ which addresses an important
issue of the spurious divergences in the nonrelativistic effective theory of
unstable top-quark production.
\section{Eliminating the spurious divergences}
\label{sec::log}
Let us first outline the problem. In the pNRQCD perturbation theory the
Coulomb Green's function gets corrections due to the second term of
eq.~(\ref{currentexp}) and from the relativistic corrections to the Coulomb
Hamiltonian
\begin{equation}
\delta{\cal H}=-{{\bfm{\partial}}^4\over4m_q^3}+{C_F\alpha_s\over2m_q^2}
\left\{{\bfm{\partial}}^2,{1\over x}\right\}\,,
\label{delham}
\end{equation}
of the following form
\begin{eqnarray}
\delta G_1(0,0;E)&=&{5\over 3}{E\over m_t}G_1(0,0;E)\,.
\label{delg1}
\end{eqnarray}
As it has been pointed out the Green's function in the above equation is
divergent and in dimensional regularization reads
\begin{eqnarray}
G^{\epsilon}_1(0,0;E)
&=&\int {{\rm d}^{d-1}\bfm{p}\over (2\pi)^{d-1}}
{{\rm d}^{d-1}\bfm{p}'\over (2\pi)^{d-1}}{m_t\over{\bfm{p}^2-m_tE}}
{m_t\over{\bfm{p'}^2-m_tE}}{4\pi C_F\alpha_s\over
(\bfm{p}-\bfm{p}')^2}
\nonumber \\
&=&-{C_F\alpha_sm_t^2\over
8\pi}\left[-{1\over 2\epsilon}+\ln\left({-m_tE\over
\mu^2}\right)-1+2\ln{2}+{\cal O}(\epsilon)\right]\,,
\label{g1dimreg}
\end{eqnarray}
where the standard $\overline{\rm MS}$ factor $\left({\mu^2e^{\gamma_E}\over
4\pi}\right)^\epsilon$ per loop is suppressed. For real energy values the
divergent part
of eq.~(\ref{delg1}) is real and does not contribute to the cross section. After
the complex energy shift the divergence gets the imaginary part proportional to
$\Gamma_t$
\begin{eqnarray}
 \lefteqn{ {\rm Im}[\delta G^{\epsilon}_1(0,0;E+i\Gamma_t)]_{\Gamma_t}
={5\over 3}{\Gamma_t\over m_t}\,{\rm Re}[G^{\epsilon}_1(0,0;E+i\Gamma_t)]}&&
\nonumber \\
&=&-{5\over 24}\left[-{1\over
2\epsilon}+\ln\left({m_t\sqrt{E^2+\Gamma_t^2}\over
\mu^2}\right)-1+2\ln{2}+{\cal O}(\epsilon)\right]{C_F\alpha_s\over\pi}m_t\Gamma_t\,
\label{delg1pot}
\end{eqnarray}
resulting in a divergent cross section. In the previous  analysis the Green's
function was renormalized, as in eq.~(\ref{g1}), leaving a finite but
scheme-dependent cross section. The solution of this problem is straightforward
within our approach. Indeed the logarithmically divergent integral in
eq.~(\ref{delg1pot}) has a physical cutoff scale $\rho^{1/2}m_t$ where the
imaginary part of the top-quark mass operator vanishes. In the above expression
this scale is set to infinity. Thus within the expansion by regions framework
eq.~(\ref{delg1pot}) represents a contribution of the pNRQCD potential momentum
region $|\bfm{p}|\sim \sqrt{m_tE}$. To get the total result one has to add the
contribution of the additional potential region $|\bfm{p}|\sim \rho^{1/2}m_t$.
The latter is given by the interference of the corrections to the Green's
function due to the top quark mass operator~(\ref{sigma0exp}) and due to the
relativistic corrections to the vector current~(\ref{currentexp}) and the
Hamiltonian~(\ref{delham}) in the second order of time-independent perturbation
theory. The resulting  integral  after expansion in $E/(\rho m_t)$ is infrared
divergent and in dimensional regularization reads
\begin{eqnarray}
\lefteqn{{\rm Im}[\delta G^{\epsilon}_1(0,0;E)]_{|\bff{p}|\sim \rho^{1/2}m_t}}&&
\nonumber \\
&=&{5\over 3}{\Gamma_t\over m_t}\,
\int {{\rm d}^{d-1}\bfm{p}\over
(2\pi)^{d-1}} {{\rm d}^{d-1}\bfm{p}'\over
(2\pi)^{d-1}}{m_t^2\over{\bfm{p}^2\bfm{p'}^2}}
{4\pi C_F\alpha_s\over (\bfm{p}-\bfm{p}')^2}\,
\theta({\rho m_t^2-\bfm{p}^2})
\left(1-{\bfm{p}^2\over \rho m_t^2}\right)^2
\nonumber \\
&=&-{5\over 24}\left[{1\over
2\epsilon}-\ln\left({\rho m_t^2\over \mu^2}\right)
+{5\over 2}+{\cal O}(\epsilon)\right]{C_F\alpha_s\over\pi}m_t\Gamma_t\,.
\label{delg1add}
\end{eqnarray}
In the sum of eqs.~(\ref{delg1pot}) and (\ref{delg1add}) the poles in $\epsilon$
cancel each other and one gets the finite result for the correction
\begin{equation}
{\rm Im}[\delta G_1(0,0;E+i\Gamma_t)]_{\Gamma_t}=-{5\over 24}\left[L_E
+{3\over 2}+2\ln 2\right]{C_F\alpha_s\over\pi}m_t\Gamma_t\,.
\label{delg1res}
\end{equation}
One formally reproduces the sum of the two regions, eq.~(\ref{delg1res}), if in
eq.~(\ref{delg1}) the $\overline{\rm MS}$  renormalized  Green's
function~(\ref{g1}) is evaluated with $\mu=e^{-5/4}\rho^{1/2} m_t$, which can
be taken as a practical prescription for the calculation of the cross
section. A similar analysis in the case of $P$-wave heavy-quarkonium
production has been performed in refs.~\cite{Bigi:1991mi,Penin:1998ik}. Note
that eq.~(\ref{delg1res}) gives only a part of the ${\cal
O}(\rho^{1/2}\alpha_s)$  corrections to the nonresonant cross section which
corresponds to  specific terms in the nonrelativistic expansion of
diagram~\ref{fig::ascorr}$(a)$ and does not account for the axial coupling of
the top quark and the higher order terms in the expansion of the mass operator
as well as the contribution of other diagrams which vanish in the lower orders.
The complete result for the  ${\cal O}(\rho^{1/2}\alpha_s)$ corrections can in
principle be obtained within the approach described in
section~\ref{sec::rhoexp}. However, as we will see in the next section, the
leading  term of the expansion in $\rho$ gives a good approximation of the total
series and is sufficient for the practical applications.
\section{Results}
\label{sec::res}
The NLO nonresonant contribution  to the threshold top-antitop production in
electron-positron annihilation reads
\begin{eqnarray}
  \lefteqn{ R^{(0)}_{nr} = - {8N_c\over \pi \rho^{1/2}}{\Gamma_t\over m_t}\,
   } &&
  \nonumber\\
  && \times \left\{ \left[ \qe^2\qt^2 + \frac{2\qe\qt\ve\vt}{1-x_Z} +
      \frac{(\ae^2+\ve^2)\vt^2}{(1-x_Z)^2} \right] f_a^{VV} - \left[
      \frac{2\qe\qt\ve\at}{1-x_Z} + \frac{2(\ae^2+\ve^2)\at\vt}{(1-x_Z)^2}
    \right] f_a^{VA} \right. \nonumber\\
  && + \frac{(\ae^2+\ve^2)\at^2}{(1-x_Z)^2} f_a^{AA} + \left[ \qe^2\qb^2 +
    \frac{2\qe\qb\ve(\ab+\vb)}{1-x_Z} +
    \frac{(\ae^2+\ve^2)(\ab+\vb)^2}{(1-x_Z)^2} \right] f_b \nonumber\\
  && + \left[ \qe^2 + \frac{2\qe\ve\cw}{\sw (1-x_Z)} +
    \frac{(\ae^2+\ve^2)\cw^2}{\sw^2(1-x_Z)^2} \right] f_c
  + \left[\frac{\qe\qb\ve\at}{1-x_Z} +
    \frac{(\ae^2+\ve^2)\at(\ab+\vb)}{(1-x_Z)^2} \right] f_d^A \nonumber\\
  && + \left[ \qe^2\qt\qb + \qe\ve \frac{\qt(\ab+\vb)+\qb\vt}{1-x_Z} +
    \frac{(\ab+\vb)(\ae^2+\ve^2)\vt}{(1-x_Z)^2} \right] f_d^V
  \nonumber\\
  && + \left[ \qe^2\qt + \qe\ve \frac{\qt\cw + \vt\sw}{\sw (1-x_Z)} +
    \frac{(\ae^2+\ve^2)\vt\cw}{\sw(1-x_Z)^2} \right] f_e^V -
  \left[ \frac{\qe\ve\at}{1-x_Z}  + \frac{(\ae^2+\ve^2)\at\cw}{\sw (1-x_Z)^2}
  \right] f_e^A \nonumber\\
  && + \left[ \qe^2\qb + \qe\ve \frac{\qb\cw+(\ab+\vb)\sw}{\sw(1-x_Z)} +
    \frac{(\ae^2+\ve^2)(\ab+\vb)\cw}{\sw(1-x_Z)^2} \right] f_f \nonumber\\
  && + \frac{1}{\sw^4} f_g + \left[ \frac{\qe\qt}{\sw^2} +
      \frac{(\ae+\ve)\vt}{\sw^2(1-x_Z)} \right] f_h^V +
    \frac{(\ae+\ve)\at}{\sw^2(1-x_Z)} f_h^A \nonumber\\
  && \left. + \left[ \frac{\qe\qb}{\sw^2} +
      \frac{(\ae+\ve)(\ab+\vb)}{\sw^2(1-x_Z)} \right] f_i + \left[
      \frac{\qe}{\sw^2} + \frac{(\ae+\ve)\cw}{\sw^3(1-x_Z)} \right] f_j
  \right\} \,, \label{eq::res}
\end{eqnarray}
where $f_n=f_n(\rho)$ stands for the contribution of
diagram~\ref{fig::dias}$(n)$ and the superscript distinguishes the contribution
of the vector and axial top-quark coupling. Each function $f_n$ is found as
a power series in $\rho$ up to $\mathcal{O}(\rho^{12})$. A text file with the
expressions for all $f_n$ is attached to the paper's source files on the arXiv.
The prefactor of eq.~(\ref{eq::res}) is chosen in such a way that $f_a^{VV} = 1
+ \mathcal{O}(\rho)$.
In order to cross-check our results we also expanded the integral
representations given in ref.~\cite{Beneke:2010mp} and found perfect
agreement.

\begin{figure}[t]
  \begin{center}
    \includegraphics[width=0.48\textwidth]{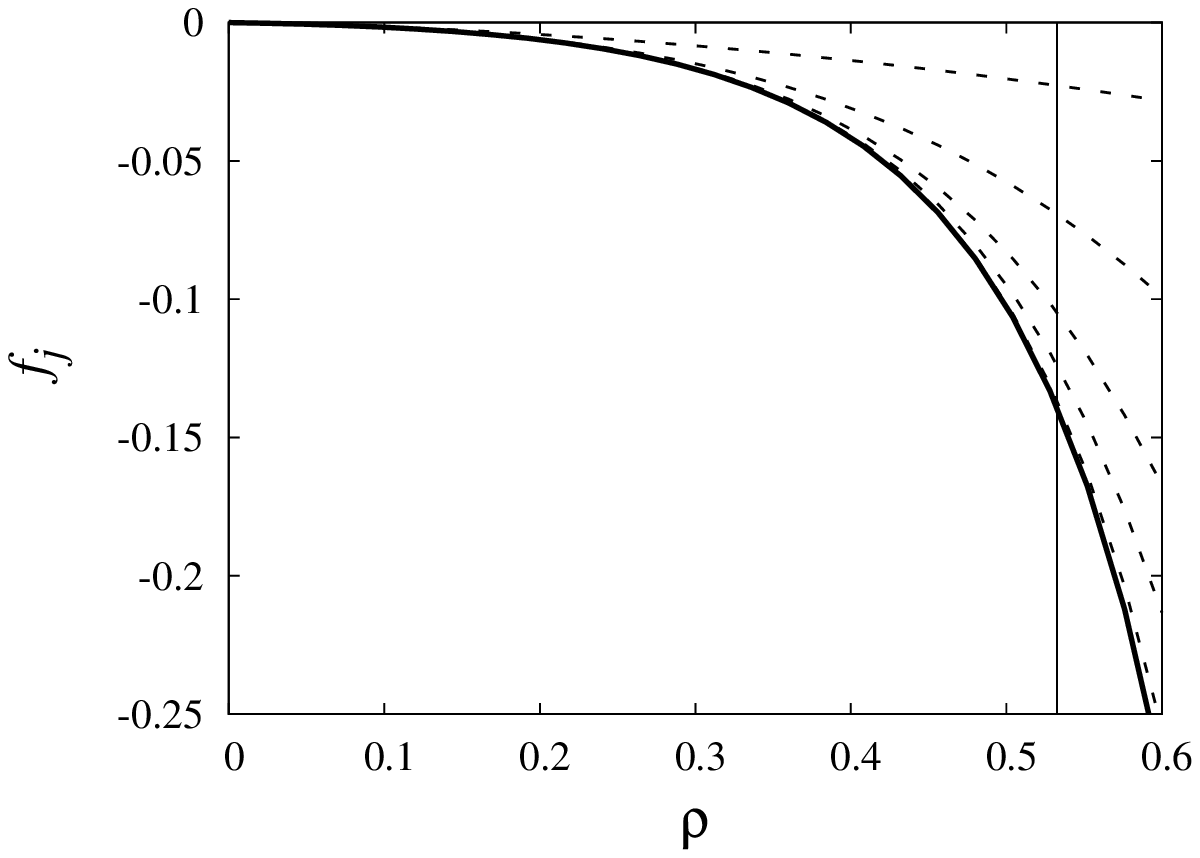}
    \hfill
    \includegraphics[width=0.48\textwidth]{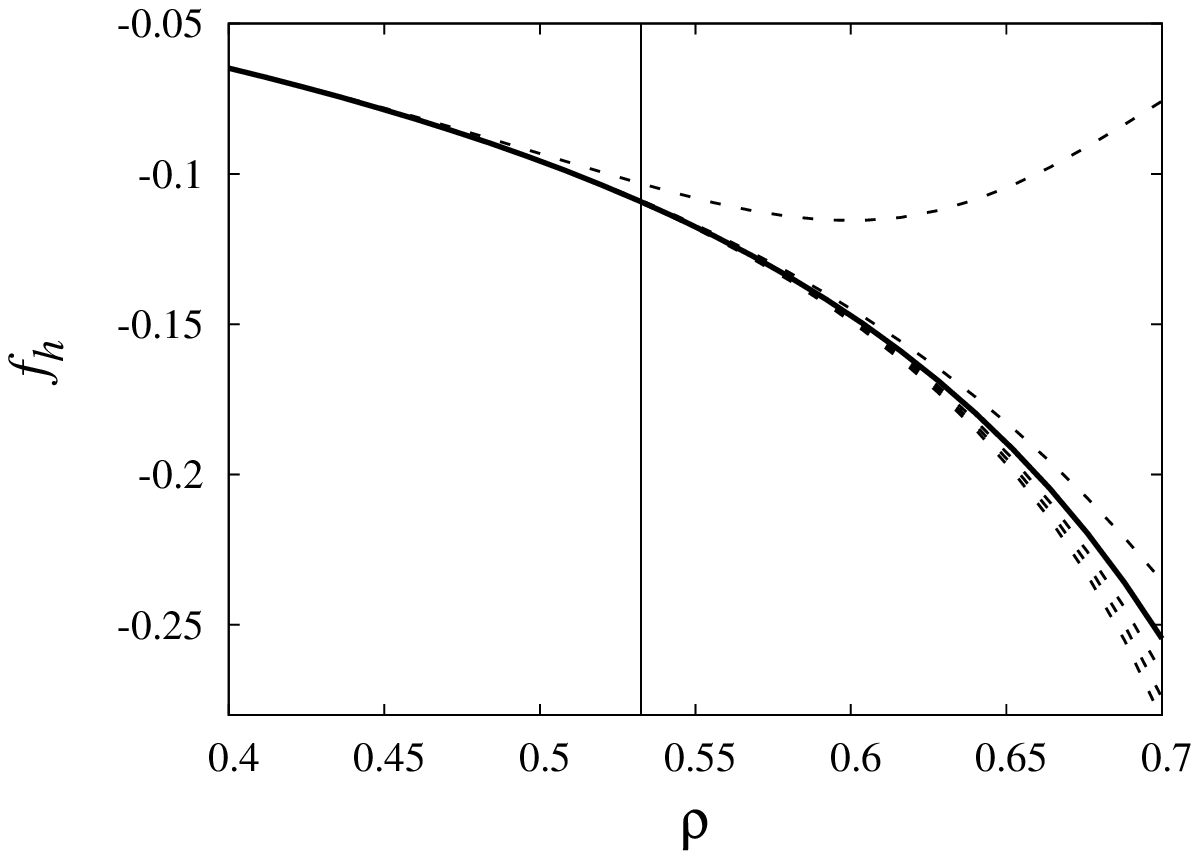}\\
\hspace*{35mm}   $(a)$ \hfill $(b)$  \hspace*{30mm}
\end{center}
\caption{\label{fig::conv} $(a)$  Dashed lines represent the expansion of
the function $f_j(\rho)$ in $\rho$ through ${\cal O}(\rho^{2n})$ for
$n=1,\ldots, 6$. $(b)$ Dashed lines represent  the $[1/11]$, $[11/1]$,
$[9/3]$, $[6/6]$, and $[3/9]$  Pad\'{e} approximants of the series for the
function $f_h^V(\rho)$.  In both pictures the vertical line marks the physical
value of $\rho$ and the solid curves represent the all-order numerical result of
ref.~\cite{Beneke:2010mp}.}
\end{figure}

With the exception of $f_h^V$ and $f_h^A$, the series
converge very well for the physical value $\rho\approx 0.53$.
As an example, in  figure~\ref{fig::conv}$(a)$  we compare
the expansion of the function $f_j$  with the exact numerical result
obtained from the integral representation given in ref.~\cite{Beneke:2010mp}.
For the two special cases we perform a Pad\'{e} resummation to improve the
convergence, i.e. we construct the Pad\'{e} approximants
\begin{equation}
  [n/m] = \frac{\sum_{i=k}^n a_i \rho^i}{1 + \sum_{i=1}^m b_i \rho^i}\,,
  \label{eq::pade}
\end{equation}
where $k$ equals $1$ and $2$ for $f_h^V$  and $f_h^A$, respectively. The
coefficients $a_i$ and $b_i$ are determined by matching the expansion of
eq.~(\ref{eq::pade}) in $\rho$  to the series for $f_h^V$ and $f_h^A$. In
figure~\ref{fig::conv}$(b)$  we compare different Pad\'{e} approximants of
$f_h^V$ with the result of ref.~\cite{Beneke:2010mp} and find perfect numerical
agreement, which means that Pad\'{e} resummation solves the problem of slow
convergence. For our numerical analysis of the cross section  we choose the
approximant $[9/3]$ for $f_h^V$ and $[8/3]$ for $f_h^A$, which deviate from the
result of ref.~\cite{Beneke:2010mp} by less than one percent for the physical
value of $\rho$.

\begin{figure}[t]
  \begin{center}
    \includegraphics[width=0.6\textwidth]{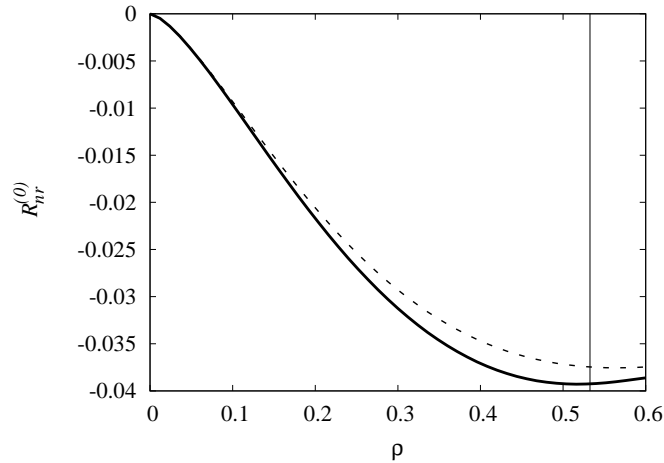}
  \end{center}
  \caption{\label{fig::dR1} The NLO nonresonant contribution to the cross
    section~(\protect\ref{eq::res}) (solid curve) and the leading order
    approximation~(\protect\ref{rnrlo}) (dashed line). The vertical line marks the
    physical value of $\rho$.}
\end{figure}
The  NNLO nonresonant contribution in the leading order in $\rho$ is given by
the ${\cal O}(\alpha_s)$ correction to eq.~(\ref{rnrlo}). By combining
eqs.~(\ref{d1acoul}), (\ref{d11ahard}), (\ref{d11b}), and (\ref{d12}) one gets
\begin{eqnarray}
R^{(1)}_{nr} &=& \frac{N_c C_F\alpha_s}{\pi^2 \rho}\,
\frac{\Gamma_t}{m_t} \bigg\{ \left[ \qe^2\qt^2 +
\frac{2\qe\qt\ve\vt}{1-x_Z} +
\frac{(\ae^2+\ve^2)\vt^2}{(1-x_Z)^2} \right]
\nonumber\\
&&\times\left[\left(3L_E+\frac{3}{2}+6\ln2\right){\pi^2}
+\left(18 + 24\ln2 \right)\rho^{1/2}\right]
\nonumber
\\
&& + \frac{1}{\sw^4} \bigg[ \frac{22}{3} + \frac{17\pi^2}{6} -
\frac{17}{2} \ln2 + \left(2 - 3\pi^2 + 9\ln2 \right) \frac{3\sqrt{2}}{4}
\ln\left(1+\sqrt{2}\right)
\nonumber
\\
&& - \frac{27\sqrt{2}}{8} \left( \ln^2\left(1+\sqrt{2}\right) +
\mathrm{Li}_2\left(2\sqrt{2}-2\right) \right) \bigg]\rho^{1/2}
+{\cal O}(\rho)\bigg\}\,.
\label{r1res}
\end{eqnarray}
Let us now study the numerical effect of the correction. We adopt the following
input values \cite{Nakamura:2010zzi}
\begin{eqnarray}
  & m_t = 172~\mathrm{GeV}\,, \qquad m_W = 80.399~\mathrm{GeV}\,, \qquad
  m_Z = 91.1876~\mathrm{GeV}\,, & \nonumber \\
  & G_F = 1.16637\cdot 10^{-5}~\mathrm{GeV}^{-2}\,, \qquad
  \alpha_s(m_Z)=0.1184 \,. &
\end{eqnarray}
In eq.~(\ref{r1res}) we use the value of the strong coupling constant
$\alpha_s(\mu) = 0.1129$ corresponding to the physical normalization scale
$\mu=\rho^{1/2}m_t$. This value is obtained from $\alpha_s(m_Z)$ by means of the
{\tt RunDec} program \cite{Chetyrkin:2000yt}.

\begin{figure}[t]
  \begin{center}
    \includegraphics[width=0.6\textwidth]{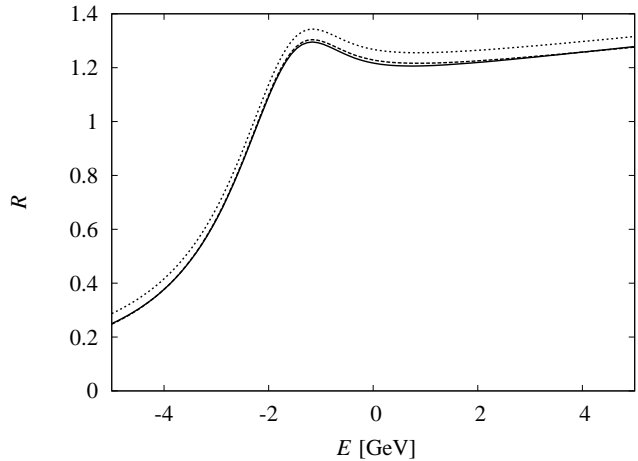}
  \end{center}
  \caption{\label{fig::full} The normalized cross section of top-antitop
  production in electron-positron annihilation as  function of the
  energy counted from the  threshold. The dotted curve represents the leading
  order pNRQCD Coulomb approximation. The dashed curve includes the
  NLO nonresonant contribution  and the solid curve includes the NNLO
  nonresonant correction as well. No strong coupling corrections to the
  resonance contribution are included.}
\end{figure}

Our total result for the NLO contribution~(\ref{eq::res}) is plotted  in
figure~\ref{fig::dR1} as function of $\rho$ along with  the leading order
term~(\ref{rnrlo}). The latter turns out to be a good approximation in the whole
interval $0<\rho<0.6$ and deviates from the total result by less than $5\%$ at
the physical value of $\rho$. This justifies our approximation of the NNLO
nonresonant contribution~(\ref{r1res}) by the leading order of the expansion in
$\rho$. The numerical effect of the nonresonant contribution on the total
threshold cross section is shown in figure~\ref{fig::full}. In this plot we use
the leading order pNRQCD approximation for the resonance contribution
corresponding to the Coulomb Green's function~(\ref{gc}) with the strong
coupling constant normalized at the soft scale $\mu_s=\alpha_s(\mu_s)C_Fm_t$.

\section{Measured cross section and invariant mass cuts}
\label{sec::cut}
\begin{figure}[t]
  \begin{center}
    \includegraphics[width=0.6\textwidth]{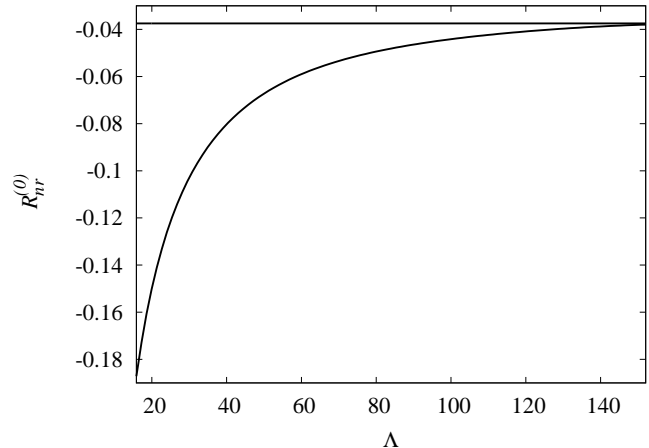}
  \end{center}
\caption{\label{fig::cut} The nonresonant contribution as function of the
cut on the three-jet final state invariant mass. The
horizontal line corresponds to eq.~(\protect\ref{rnrlo}).}
\end{figure}

Experimentally the top-antitop pairs produced in electron-positron annihilation
are to be reconstructed from the lepton plus four jets final state or all
hadronic six jet events. Realistic simulations show that the invariant mass
$p^2=(p_W+p_b)^2$ of the three jets resulting from the bottom quark and the
hadronic decay of the $W$-boson can be determined and the corresponding
invariant mass distribution can be obtained with high accuracy
\cite{Chekanov:2003cp}. The total cross section of the threshold production
considered in this paper is obtained by integrating this distribution over the
kinematically allowed interval $m_W^2<p^2<m_t^2$, which corresponds to the
integral over $\bfm{p}$ in eqs.~(\ref{d10}) and (\ref{d20}). One may suggest
that a tight invariant mass cut $m_t^2-p^2<\Lambda^2$ with  $\Lambda\sim
(m_t\Gamma_t)^{1/2}$ would separate the  production of ``true'' on-shell top
quark and antiquark from the nonresonant background. However, it is not possible
to suppress the nonresonant contribution without significant modification of the
resonant one. The resonant contribution of unstable top quark corresponds to the
Breit-Wigner shape of the invariant mass distribution, which falls off rather
slowly as the invariant mass deviates from $m_t$  and is
strongly affected even by loose cuts. To quantify this statement let us consider
a loose cut $(m_t\Gamma_t)^{1/2}\ll\Lambda$.  In the Born approximation it can
be implemented by replacing the physical cutoff $\rho m_t^2$ in the argument of
the theta-functions in eqs.~(\ref{d10}) and~(\ref{d20}) with $\Lambda^2/2$. For
$\Lambda\ll \rho^{1/2}m_t$ one gets
\begin{equation}
\delta^{(0)}_1=-{3\sqrt{2}\over \pi}{m_t\over \Lambda}
+{\cal O}\left({\Lambda\over\rho^{1/2}m_t}\right)
\label{d10Lam}
\end{equation}
and  $\delta^{(0)}_2\sim (\Lambda/\rho^{1/2}m_t)^3$. The leading term in
eq.~(\ref{d10Lam}) corresponds to the first term of eq.~(\ref{d10}) and
represents the modification of the resonant contribution to the cross
section~(\ref{res}) by the cut. It is merely the dominant effect of the cut
unless $\Lambda\sim \rho^{1/2}m_t$. The dependence of the correction to the
cross section on the invariant mass cut value is shown in
figure~\ref{fig::cut}. Starting from $\Lambda\sim 80$~GeV  it is relatively
weak and the correction is well approximated by eq.~(\ref{rnrlo}).
This gives a low bound on the acceptable cut value which can be used for the
determination of the total cross section. One may study the cross section with
tighter cuts as well. However all the high order QCD results for the total cross
section are not applicable in this case and accuracy of the theoretical
predictions  would be significantly reduced.
\section{Summary}
\label{sec::con}
We have developed a new method for the analysis of the top-quark
instability in the threshold top quark-antiquark pair production
in electron-positron annihilation based on the nonrelativistic
expansion in the  parameter $\rho=1-m_W/m_t$. Within this framework
we obtain the  NLO  nonresonant contribution to the total threshold
cross section overlooked in the standard analysis and confirm the
result of ref.~\cite{Beneke:2010mp} obtained within a different
effective theory approach without the expansion in $\rho$.  The NLO
contribution  is negative and amounts to about 3.1\% of the leading
cross section above the threshold and competes with the LO contribution
below the resonance region. We extend the analysis to the NNLO
${\cal O}(\alpha_s)$ nonresonant contribution which is computed to the
leading order  in $\rho$. The corrections involve $\ln\rho$ terms
which are a new type of the large logarithms in the theory of top
quark threshold production. The NNLO nonresonant contribution amounts
to $-0.9\%$ at the threshold.

Our method can also be applied to the calculation of the nonresonant
contribution to the threshold top quark-antiquark pair  production
in hadronic collisions. The process is of particular interest since a
significant number of top quark-antiquark pairs is going to be produced at the
LHC near the threshold and the accuracy of the top quark reconstruction is
expected to be sufficiently good to study the threshold region. A comprehensive 
analysis of the process is given in refs.~\cite{Hagiwara:2008df,Kiyo:2008bv},
where the NLO approximation without the  nonresonant part has been used to
derive the  top-antitop invariant mass distribution. The missing nonresonant
contribution can be directly obtained from our result. Indeed, in the threshold
region the cross section is dominated by the color singlet top-antitop
configuration which is produced mainly through gluon fusion. Thus, the dominant
nonresonant contribution  is entirely due to the  imaginary part of the
top-quark mass operator and in full analogy with  eqs.~(\ref{res})
and~(\ref{r1}) one gets
\begin{equation}
R_{nr}({\rm hadrons}\to t\bar{t})=
{16\over 3\pi}\left({2\Gamma_t\over \rho m_t}\right)^{1\over 2}
\left. R^{Born}_{res}({\rm hadrons}\to t\bar{t})\right|_{E=i\Gamma_t}
(1+{\cal O}(\rho,\alpha_s))\,,
\end{equation}
which should be added to the resonant contribution considered in  
refs.~\cite{Hagiwara:2008df,Kiyo:2008bv}.

\acknowledgments
We are grateful to A. Czarnecki, K. Melnikov, and A. Pak for useful
communications, and M. Beneke and M. Steinhauser for carefully reading the
manuscript and useful comments. JHP thanks B. Jantzen and P. Ruiz-Femen\'{i}a
for helpful conversations about ref.~\cite{Beneke:2010mp} and the Institut
f{\"u}r Theoretische Teilchenphysik at the Karlsruhe Institute of Technology for
hospitality. This work was supported  by NSERC, the Alberta Ingenuity Foundation
and the DFG Sonderforschungsbereich/Transregio 9 ``Computergest{\"u}tzte
Theoretische Teilchenphysik''. The work of AP is supported in part by Mercator
DFG grant. The Feynman diagrams were drawn with {\tt JaxoDraw
2}~\cite{Binosi:2008ig}.

\end{document}